# Multi-Omics Reveals Temporal Scales of Carbon Metabolism in *Synechococcus elongatus* PCC 7942 Under Light Disturbance

Connah G. M. Johnson[1†], Zachary Johnson[2,3†], Liam S. Mackey[4†], Xiaolu Li[2], Natalie C. Sadler[2], Tong Zhang[2], Wei-Jun Qian[2], Pavlo Bohutskyi[2,3], Song Feng[2*], Margaret S. Cheung[2, 5, 6*],

**1** Physical and Computational Sciences Directorate, Pacific Northwest National Laboratory, Richland, Washington, 99352, United States of America

**2** Biological Sciences Division, Pacific Northwest National Laboratory, Richland, Washington, 99352, United States of America

**3** Washington State University, Pullman, Washington, 99164, United States of America

**4** Rensselaer Polytechnic Institute, Troy, New York, 12180, United States of America

**5** Environmental Molecular Sciences Laboratory, Pacific Northwest National Laboratory, Richland, Washington, 99354, United States of America

**6** University of Washington, Seattle, Washington, 98195, United States of America

* Co-corresponding authors

Margaret S. Cheung, margaret.cheung@pnnl.gov, Song Feng, song.feng@pnnl.gov,

†These authors contributed equally to this work.

**Author contributions**

M.S.C., S.F., W.Q., T.Z., P.B. designed the research. C.J., Z.J., L.M., S.F., X.L., N.S. performed the research. M.S.C., C.J., Z.J., X.L., S.F. analyzed data. All authors contributed to writing the paper.

**Competing Interest Statement: No competing interest**

## Abstract

Central carbon metabolism in model cyanobacteria involves multiple pathways to adapt to energy-light limitations across diel cycles. However, the success in mechanistic modeling for phenotypic prediction of the protein regulators in the metabolic state depends on capturing the vast possibilities emerging from multiple regulatory pathways in complex biological processes. Here, we developed a physics-informed machine learning approach based on energy-landscape concepts to predict regulatory proteins responding to cyclic circadian and unforeseen light perturbations in cyanobacterial metabolic networks. Our approach provides interpretable de novo models for inferring gene expression dynamics from *Synechococcus elongatus* over diel cycles and using redox proteome analysis to distinguish immediate light-responsive elements from circadian-regulated processes in carbon metabolism pathways. We identified distinct temporal signatures with the analysis of the redox proteome: there was an immediate shift in cysteine redox states accompanied by a limited change in protein abundance under constant illumination and after two hours of darkness. This discovery indicated that the generation of reductants coordinates photo-induced electron transport with redox metabolic pathways in two discernable molecular mechanisms: there are fast redox-based protein modifications immediately after the light disturbance, followed by slow transcriptional regulations across networks. This temporal regulation reveals how metabolic networks integrate rapid light responses with programmed circadian rhythms to maintain cellular homeostasis under the light-energy limitations over the diel cycle.

## Significance Statement

We constructed a system-level approach to uncover the relationships between molecular and phenotypic changes under light disturbance. We used the publicly available gene expression data of cyanobacteria to train interpretable machine-learning models that generate testable, de novo networks, for elucidating complex reactions in the dynamic regulatory pathways of central carbon metabolism. This model leads to the discovery of protein regulators analyzed by redox proteome analysis. The assemblage of protein regulators directs multiple pathways in accordance with programmed circadian rhythms, offering insights into cyanobacteria's response to light or energy limitations over the diel cycle.

# Introduction

As phototrophic organisms, cyanobacteria harvest energy from light to fix $CO_2$ into glucose which can then be converted to valuable chemical commodities while eliminating the need for sugar feedstocks (1). However, the inefficiency of the enzymes in the $CO_2$ fixation pathway limits their utility as a host for biological chemical production (2). Such a shortcoming motivates the strategies for profiling alternative pathways or using synthetic biology tools to rewire metabolic processes involved in photosynthesis, $CO_2$ fixation, and carbon metabolism (3). To address this need, we used the cyanobacteria *Synechococcus elongatus* PCC 7942 (*S. elongatus)* as a tractable model for developing a data-driven modeling with a perturbation biology approach to gain insight about the metabolic processes from multi-omics data.

The advent of multi-omic technologies has ushered in the generation of vast and diverse biomolecular data for profiling biological systems (4). These multi-omics datasets were often produced within isolated contexts (5), which complicates their integration for mechanistic insights. Moreover, a common approach to modeling metabolic processes at the subcellular level (6-8) involves characterizing the transcriptomic changes induced by systematic treatments or environmental perturbations (9). These orchestrated response of a cell to stimuli is often regulated by a limited number of transcriptional factors (TFs) that modulate transcription (10-12) through conformational changes in proteins and nucleic acids. Some of these TFs are proteins or enzymes susceptible to post-translational modifications (PTMs) (13), including environmental redox responses via reactive cysteine residue (14). These PTMs on proteomes further complicate the prediction performance of systems modeling (15).

Such complex processes necessitate the development of nonequilibrium physics frameworks and data-driven machine-learning approaches to dissect the underlying mechanisms of emergent phenomena across scales (16). On a general framework for discussing reaction dynamics in complex systems (17), new phenomena can rise from the mesoscopic dynamics of intrabasin or interbasin motion on a non-equilibrium energy landscape (**Figure 1A**). There are fast, microscopic dynamics representing reversable redox reactions (18) happening at mixing at a small scale; intrabasin motions within protein conformations (19) at an intermediate scale; and the slow interbasin switching with the kinetic rates exponentially dependent on the system size at a larger scale for biochemical regulations. As noted by Anderson's principle of "More is Different", large-scale behavior emerging from microscopic scales can significantly differ from the small-scale behavior itself (20). We use light disturbance as a perturbation approach to differentiate the impact of redox interactions on the protein conformations from the protein-DNA regulatory interactions in multi-layered regulatory networks in **Figure 1B**.

In this work, we developed a physics-informed machine learning (PIML) (21) approach, based on the thermodynamics of perturbation decay towards system-wide steady states (22), into an expanded workflow to guide experimental insights. We use this approach to learn from the response of TF proteins in changing environments for predicting multiple phenotypes or profiling alternative pathways in regulatory functions. This is achieved by first applying dimension reduction techniques on the large datasets to mitigate computational demand with machine-learning method that incorporates a physical perspective (23), while recapitulating the selected phenotypes responsive to light stimuli as perturbation. To explore a broad parameter space covering a dynamical system of *S. elongatus* periodically facing light-energy limitation, we analyzed the transcriptome expression profiles under varying light intensities over the circadian rhythms in diel cycles from the public National Center for Biotechnology Information (NCBI) database. The PIML approach (**Figure 1B)** constructs dynamic models with optimal gene interaction parameters in response to light disturbance without relying on *a priori* information about their interaction network. This workflow also takes the redox and global proteome analysis on *S. elongatus* culture after light disturbance to screen for the protein regulars. The redox proteome analysis on the *S. elongatus* culture revealed that the cysteines from four protein regulators showed significant shifts in the oxidation changes, while the relative protein abundance remained similar.

These screened regulator proteins are involved in the circadian cycle, oxidative pentose phosphate pathway, and the Calvin-Benson cycle (**Figure 1B**), forming a "timed shunt" for directing biochemical reactions in multiple pathways at the protein structure level through fast redox PTM to support the metabolism of the cyanobacteria model in day-night cycles. Our work highlighted a multilayered regulatory network of central carbon metabolism built on chemical and molecular interactions, coupled to programmed circadian rhythmicity in cyanobacteria, crucial for evolutionary fitness under the periodic light-energy limitations observed in diel cycles. Our approach to integrate -omics data to screen for regulators is organism-agonistic as long as the datasets were accumulated on the basis of a perturbation biology approach.

## Results

## Development of an iterative workflow to curate metadata sets by learning dependent variables in the expression profiles responsive to stimuli over diel cycle

We established an iterative workflow of data analysis that utilizes a physics-informed machine-learning (PIML) package, CellBox (24), to curate a set of transcriptome expression data that recapitulates the selected phenotypes responsive to light stimuli as perturbation in **Figure 2**. This package relies on quantitative datasets that observe molecular changes to construct the mathematical representations of dynamical equations for systems modeling and does not require *a priori* knowledge of known molecular interactions to construct a machine-learning model. We used the expression profiles from transcriptomes of the *Synechococcus elongatus* PCC 7942 cyanobacteria (25-28) as phenotypic observables in response to light stimuli in the PIML modeling with CellBox.

The first step of the workflow is to curate a dataset of metadata filtered from the full transcriptome database of *S. elongatus* by a quality control (QC) pipeline for the phenotypic observables in response to light in a dynamic system periodically facing energy-light limitations. Please see **Section S1** in **Supplementary Information** for a description of the QC pipeline that produces comprehensive high-quality RNA-sequencing dataset from the literature including experiments on *Synechococcus elongatus* PCC 7942 circadian system. See the raw RNA-sequencing datasets in **Table S1** from the Supplementary Spreadsheet and the one passing QC in **Table S2** of the **Supplementary Spreadsheet,** respectively. Next, we transformed the expression profiles of transcriptome dynamics, quantitative observables, and dependent variables over a diel cycle to independent variables (see **Tables S3-S10** and **Section S2** in the **Supplementary Information).**

The curated dataset remains a very high-dimensional gene-expression space, while the phenotypic space perturbed by illumination resides on a much lower-dimensional manifold. We cannot assume that we fully understand the nature of the laws that govern the cell behavior (29). It remains unclear whether individual "marker" genes or groups of functionally related genes (i.e., clusters) more faithfully capture phenotypic variation, and how these gene clusters interact to form a coarse-grained description of cell states. To address the question of how to understand large and complex systems that operate across many orders of magnitude in space and time, where fluctuations at small scales may not average out but instead may be passed to higher scales to drive emergent behaviors of the whole system, we analyzed the transcriptomes of *Synechococcus elongatus* <u>after the QC pipeline</u> at two levels of granularity: 1) **Gene-level model (or Gene model)**, a fine-grained representation in which each observable corresponds to the expression level of a single gene, providing maximal resolution of transcriptomic changes; 2) **iModulon-** **level model (or iModulon model)**, a coarse-grained representation derived by applying independent component analysis (ICA) (30) to the whole transcriptome data of *S. elongatus* PCC 7942 after the QC pipeline processing, yielding “iModulons”—modules of co-regulated genes that collectively define lower-dimensional regulatory programs. Using the ICA, we derived 63 independently regulated gene sets (iModulons) (9) (see **Figure S2** and **S3**).

Both Gene and iModulon models were then used as inputs to separately train the mathematical representations of dynamical response by light stimuli with PIML using Cellbox. We used the genes known from the literature to be circadian clock regulated (31) to train a non-linear mathematical representation of the governing dynamical equations (**Eq 1**) in the machine learning models. **Table S11** and **Section S5** show the homology match between transcription domains and its regulated gene responsible for programmed circadian rhythm adjustments in cyanobacterium *S. elongatus* PCC 7942. We iterated the step of metadata curation step by updating the filters for sample QC. We used the interpretable outcome as feedback to adjust the sample QC of metadata curation enabling us to gain insight into the contributions of significant elements and interactions between nodes in the resultant network. We iterated and evaluated the learning performance with the Pearson’s correlation until the criteria of training were met -- that is when the analysis of the predicted responses returned a validation loss that is less than 100 from the CellBox package. The validation loss in the machine-learning procedure was defined as the Euclidean distance between the physics-informed response model and the experimental reference data. The details of training system-level PIML models with the Cellbox package is provided in **Section S6** of the **Supplementary information**.

We noticed that, concurrently, there is another study from the Paulson group that independently identified similar iModulons in *S. elongatus* PCC 7942 (32). We compared the two analyses between our 63 ICA-derived iModulons and theirs reported in that publication. Of these, 41 modules (≈65 %), typically with large gene sets, are shared between the two analyses (**Figure S3)**. This close correspondence demonstrates that the core iModulon structures are highly conserved across independent workflows, lending strong support to our ICA-based regulatory modules. The fine divergences between the two iModulon analyses highlight that the differences in the training datasets can act as a distinctive fingerprint in the gene expression varied by perturbation tailored to generate specific hypothesis for experimental validation.

## The PIML procedure successfully trains the mathematical representations from quantitative datasets for node interactions in response to light disturbance

We showed that the trained models have successfully learned from the quantitative sets in the procedure from Cellbox. Details are provided in **Section S7** of the **Supplementary Information**. The best 1000 models from CellBox simulations were compiled to calculate the t-value via a one-sided t-test where the goal was to define interactions that were significantly different than 0. To define significant node interactions, a one-sided t-test was used to query significance in the learnt interaction strengths when compared to the null hypothesis of no interaction. The t-test was performed on the ensemble of models to provide inference confidence in the absence of readily applicable benchmarks. We evaluated the training procedure from CellBox based on the curated data sets with the model learning statistics in **Table S12**. When compared to the phenotypic responses to light in the data (**Figure S1)**, both Gene and iModulon models provided high Pearson correlation statistics of 0.82, demonstrating a successful training of the data-set perturbations from the procedure.

As shown in **Table S12**, the best validation loss (i.e. lowest loss during training) candidate for each parent model was found to be 17.0 for the iModulon model, with a best testing loss (i.e. loss after testing on fully unseen data) of 412.4 and the best validation loss was 7.6 for the Gene model, with a testing loss of 52.6. These low loss values during the response model training suggests that the parameters were learnt accurately to the reference data. The average correlations between nodes, however, is less significant (~0.04),

The Gene model exhibited a testing loss that was 10 times less than that of the iModulon model when comparing the learnt and training expression values. However, the former had a lower number of 866 successful candidates whose validation loss met an initial cut-off threshold of 1000. This suggests that the larger data dimension of the Gene model required a greater number of network nodes to train and was more difficult to train. The larger data dimension provided more flexibility in the parameter search landscape at the cost of an increased computation time signified by fewer successful candidates. Although the iModulon model provided a greater number of 4762 successful candidates it resulted in larger validation and testing losses. In summary, the Gene model returned fewer candidates due to increased model complexity that grows with the number of nodes training on a large dataset of genes. The candidates for the smaller dataset in the iModulon model based on the groups of genes that share regulatory functions were more readily trainable, indicating that the iModulon and Gene models differed in the accuracy and training time.

## The interpretable PIML de novo network models capture directed connections to known regulators stimulated by light

A set of ordinary differential equations with a nonlinear envelope was used to simulate the dynamic response with CellBox as a PIML tool that learned the parameters of interaction strengths between a de novo network of transcription factors and genes or iModulons interactions. Because of the interpretable outcomes from PIML, a great extent of regulatory dynamics connecting to prior knowledge about the pathways responding to light stimuli from an informative set of curated transcriptome expressions over the diel cycle was revealed by comparing the nodes with a set of known regulatory genes from the literature (i.e. regulons, **Table S11**).

We illustrated *de novo* multilayered networks in **Figure 3** from the best training models in **Table S12.** The edges between nodes show directed interactions learned from training in response to light perturbation where blue edges signify known regulatory interactions **(Table S11)**. We colored edges gray for those that are potentially new connections to other regulatory pathways coupled to light perturbation. The networking architecture allows the propagation of the response over the independently regulated gene set modules in the iModulon model (**Figure 3A**). Comparatively, in the Gene model the light response propagates through genes as shown in **Figure 3B**. The former provides a "coarse-grained" view of the interactions for interpretation, while the latter is rather "fine-grained". Below, we showed that both models can capture the literature-provided interactions. In the iModulon model, there are 102 observed nodes, of which 40 were TF expression and 62 were iModulon response. In the Gene model, there are 886 observed nodes, of which 40 were TF expression and 846 were gene expression. We have selected the top-ranked nodes, in terms of significant expression, in the de novo networks for further analysis of relational interactions.

In **Figure 3A**, most iModulons are of known experimental importance, or "regulons". Indeed, the RpaA, RpaC and competence iModulons are accounted for among the top explained variances from the ICA analysis (**Figure S2**) as well as shown in Palsson's work(32). These connected nodes in a directed graph reveal a simplified regulatory network with key members. The phenotypic RpaA iModulon (enriched in RpaA and SigF2 targets), correctly connects with the two known circadian regulators genes, RpaA and SigF2 (**Figure S2B**). RpaA is also identified as the regulator of RpaC iModulon which is enriched in RpaA and RpaB targets (**Table S11**). We demonstrated that the directed graph of the *de novo* network captures both known relational interaction between sigma factors, and downstream gene set modules (iModulons). Regulation of significant iModulons is identified to be controlled by RpaA in coordination with sigma factors. Together these regulators modulate circadian rhythm in *S. elongatus* through an interwoven web of direct and indirect relationships (28). The mathematical models capture the perturbation effects from light stimuli, even though these iModulons are not in direct relation to the light source.

The Gene model (**Figure 3B**) reveals a greater number of nodes than that for the iModulon model (**Figure 3A**), as the former represents the nodes at a more granular detail about the direct interactions than the latter. Among the known genes in the interaction networks that are responsive to light stimuli in **Figure 3B**, the RpoD5 sigma factor connects to two conserved hypothetical proteins (Synpcc7942_1397, 0316). However, the fine-grained view lacks the representation of higher-order connections through the propagation of direct interactions because the interactions between entities are comparably weak. This is manifested by the histogram of the Pearson correlation coefficients between the predicted and observed expressions for each node in the *de novo* networks in **Figure 3 (C)** and **(D)** for the iModulon and Gene models, respectively. Although the correlation coefficients are higher for the nodes from both *de novo* networks than those representing the known TFs alone (TF, in green), the distribution of correlation coefficients is broader for individual genes (**Figure 3D**) than that for iModulons in **Figure 3C**, inferring a weaker parameter interaction for the former. As a reference, a nodal correlation coefficient of 1 is shown for the perturbation nodes (in purple).

## The interpretable PIML de novo network infers interactions that connect known regulators stimulated by light to multiple pathways including central carbon metabolism

The *de novo* networks inferred from transcriptome datasets with PIML models reveal far more statistically significant relational interactions (gray edges in **Figure 3**) than known direct or indirect interactions responsive to light stimuli over the circadian time. The constructed *de novo* networks of dynamic models discover new connections (shown by gray edges) from known regulators (red nodes), inferring that the propagated gene expressions are responsive to light perturbation. A coarse-grained view of the network in the iModulon model in **Figure 3A** shows that the responsive nodes are involved in the circadian regulation as well as metabolic pathways. A set of TFs identified in the RpaA iModulon are simultaneously connected to its regulation including SrrB and putative transcriptional regulator TetR. The SrrB regulator alongside the sigma factor RpoD6, and the circadian clock protein kinase KaiC are connected to regulation of the RpaC iModulon. Additionally, the competence iModulon is connected to sigma factors RpoD6 and RpoD5 (or namely “sigC"). Interestingly, all three major iModulons; RpaA, RpaC and competence are connected to TetR. While TetR is not directly responsive to other genes under light perturbation in **Figure 3B**, the interpretation from the analysis signifies multilayered networks where TetR possibly plays a pivotal role in regulating multiple pathways. Two look-up Tables for the descriptions of the node labels in **Figure 3A** are provided **in Tables S13** and **S14.**

In the fine-grained Gene model, the strongest edges in the *de novo* network infer interactions between regulators upstream of RpoD5 and downstream gene targets. These inferred interactions implicate RpoD5 in regulating genes involved in redox metabolism, generation of reducing equivalents, and response to stress (Additional look-up table for **Figure 3B** in **Table S15)**. Notably, pyridine nucleotide transhydrogenase (*pntA*, *pntA2*, *pntB*) facilitates the conversion of NADH to NADPH, while ferredoxin-plastoquinone oxidoreductase subunit D1 (*ndhD1*) and cytochrome-c oxidase (*ctaC*) participate in oxidative phosphorylation. Additional genes, including *fbp*, *zwf*, and *opcA*, contribute to glycolysis and the oxidative pentose phosphate pathway (**Figure S4**). Alternatively, genes such as *katG*, *dnaJ*, and *uspA* have functional roles in stress response, *katG* encoding a catalase/peroxidase responsible for reduction of hydrogen peroxide.

Gene sets identified in both the iModulon and Gene models correspond to the gene sets expressed during circadian night. The RpoD5 sigma factor and predicted downstream gene targets themselves members of the RpaA iModulon (**Table S13**). Although their biochemical reactions are not directly initiated by light, their regulations that happen in cytosol may be promptly impacted by reactive oxygen species (ROS), products of photosynthetic reactions, that give rise to multiple phenotypes represented by cellular redox homeostasis in photosynthetic organisms (33). *This information connecting light stimuli to metabolic genes over circadian time allows us to test a hypothesis that redox post-translational modifications (PTM) influence the activities of proteins that regulate overlapping pathways from the central carbon metabolism over the diel cycle.*

## The redox proteomes narrow screening for protein regulators in central carbon metabolic pathways of *S. elongatus* revealed by light disturbance

The connectivity interaction, $\omega$, from the de novo kinetic model in **Figure 3** infer the interactions of gene expressions of *S. Elongatus*, or regulations, under light perturbation over circadian time, offering insights into a testable hypothesis about changing the metabolic redox state *in vivo* under light perturbation by using proteomic-based approaches. We focused on redox thiol PTMs on cysteines mediated by reactive oxygen species (ROS) that constitute a major mechanism during photosynthesis (34). We hypothesize that because it is fast for redox PTM to respond to ROS changes than needing enzymes for catalysis, we will capture protein regulators with strong changes in the redox PTM expression while their changes in protein abundance expression remain minimal.

We monitored the cellular redox state of *S. elongatus in vivo* using a mass spectrometry approach (see Method Section) aimed at investigating the dynamics of protein abundance changes as well as redox proteome modifications after light/dark stimuli. We cultured *S. elongatus* under constant light followed by a short-term (i.e. 2-hour) deprivation of light as perturbation. We first profiled the changes in the cysteine oxidation level from the redox proteome and the changes in the protein abundance from the global proteome analysis. We compare their log2 Fold Change (log2FC) values from redox proteome to the strong connectivity interactions meeting t-test significance criteria from the de novo kinetic model of gene expressions for transcription factors (TFs) in **Figure 4**. **Figure 4A** shows the magnitude of interaction strengths meeting t-test significance criteria for the transcription factors (TFs) that contribute towards the response to the light perturbations. This demonstrates the TFs such as sigC (RpoD5), SigI and SrrB drive the iModulon model response while gdh, RpoD6, and ntcA are strong drivers for the gene model response. These predictions are accompanied by the log2FC changes in the abundance expression as well as the redox expression from the redox proteome of *S. Elongatu*s under illumination perturbation (**Figure 4B**). It shows that TFs that respond to light perturbation react differently. For example, TetR is seen as significant in the de novo kinetic models but it is not significantly detected in proteomics experiments under light perturbations. For SigI, both protein abundance and redox expression change. For the gene products of pex, hup, and himA, only protein abundance expressions change, but not their redox expressions. Most interestingly, for the gene product of rpaA and dprA, their redox proteome expression changes, but not the protein abundance expression.

Under our hypothesis, we further followed the proteome of the gene products from rpaA and dprA. rpaA and dprA belong to the RpaA iModulon and competence iModulon in **Figure S5 and Figure S6**, respectively. RpaA iModulon, one of the largest iModulons in the study, include a large number of metabolic genes indirectly connected to light disturbance. Our redox proteomic approach allows us to profile the changes in the oxidation of cysteines at a residual level while monitoring the protein abundance after the light disturbance. To visualize these complex data into an intuitive figure, we organized these active proteins against the gene products inferred by the *de novo* models from the transcriptome data in a word cloud format (**Figure 5A**). The redox changes in RpaA, Fbp, Zwf, and OpcA are significant, but their changes in the protein abundance are not (cutoff: p-value $< 0.05$, log2FC $> 0.1$ or $< -0.1$) (**Figure 5B**). The oxidation level at multiple Cys sites of these proteins changed significantly by light disturbance, suggesting a rapid signaling mechanism by redox PTM. Intriguingly, the expression of cysteine PTM on the gene product of ftp, a Paprotein regulator in the Calvin-Benson cycle, became more oxidized (red) as the system changed from light to dark. In contrast, the expression of cysteine PTM on the gene product of *zwf*, a protein regulatory in the oxidative pentose phosphate pathway (OxPPP), became more reduced (blue). The cysteine-rich OpcA showed both an increase and decrease in cysteine PTM expression, indicative of the constitutive role of the *opcA* gene, a "class 2" circadian regulator, in shunting multiple pathways in the central carbon metabolism for biochemical processes in response to programmed circadian rhythmicity under limited light energy.

We also discovered few proteins in which the redox PTM and abundance changed significantly. The function of these proteins is to interact with DNAs directly such as PriA, a DNA helicase. Therefore, we do not take them into account when testing our hypothesis. In summary, the proteomic experiments, and *de novo* models from PIML were used in a novel way to infer the mechanisms that influence multiple pathways from central carbon metabolism in response to light disturbance.

# Discussion

## The PIML workflow offers interpretable de novo network models from curated transcriptomic expression datasets

Accurate regulatory network inference in cyanobacteria using conventional methodologies begins with fundamental challenges in correctly identifying transcription factors. Unlike model heterotrophs with relatively well-characterized regulatory proteins, cyanobacterial genomes contain numerous putative regulators with domain structures that diverge from canonical transcription factor families, making their identification inherently uncertain even with advanced prediction tools. Our recent work demonstrated substantial discrepancies between computational methods, with only 28% agreement between three state-of-the-art TF prediction algorithms(35). This uncertainty propagates through network inference, where even top-performing ML algorithms like GENIE3 consistently achieve modest accuracy (F1-scores of ~0.11) for direct transcription factor-gene interaction prediction. Co-expression approaches like WGCNA can identify correlated gene modules but lack the ability to determine causality or directedness in regulatory relationships. Furthermore, cyanobacteria's complex light-dependent metabolism and intertwined circadian control mechanisms introduce additional layers of regulation beyond conventional transcriptional control, including DNA topology changes and post-translational modifications that standard network inference methods cannot capture.

Our physics-informed machine learning (PIML) approach addresses these limitations by focusing on system-level dynamics and incorporating biophysical constraints that reflect the mechanistic basis of gene expression responses to perturbations. Rather than solely predicting individual interactions, PIML reveals emergent network properties that illuminate regulatory principles governing temporal metabolic transitions. This approach enables discovery of previously uncharacterized regulatory mechanisms that coordinate diurnal metabolic shifts—insights inaccessible through conventional network inference techniques. By integrating physics-based constraints with machine learning, PIML provides a complementary framework that captures the multilayered nature of regulatory networks in photosynthetic organisms, offering new perspectives on complex cellular processes that traditional inference methods often overlook.

We overcame the challenges in the implementation of PIML to learn about gene regulation from known data by adding an iterative layer of CellBox to the workflow. CellBox was first designed to interpret molecular variations from omics datasets by attributing variations to specific and well-defined phenotypes, such as those used for drug screening in cancer research (24). Microbial transcriptomic expressions are not direct applications due to: (1) the database encompasses multiple phenotypes, and (2) they often lack information of time-series responses to systematic perturbations. Moreover, it is challenging to establish a mathematical physics model with relational interactions based solely on gene expressions to explain the meaning of observations in a biological process when these sequencing experiments offer sparse transcriptome datasets. Here, we used three strategies to overcome these deficiencies: (1) We leveraged a full range of transcriptomic datasets of *S. elongatus* cyanobacteria from published databases as a tractable model to develop our workflow that includes a Quality Control (QC) pipeline, ensuring that we covered meaningful gene expressions to the datasets. (2) We remedied the sparsity of the transcriptome dataset by using the iModulon [8] to reduce the data dimensions by grouping genes into functional clusters (**Figure S2**) after the QC pipeline. (3) We followed light-dependent reactions such as photosynthesis as dependent variables to set up light stimuli as perturbations to track the propagation of events. Since these reactions are well-documented for studying this cyanobacteria's programmed circadian cycle that offers time-related information, we aggregated these gene expressions as phenotypic observations over a diel cycle to train the physics model with CellBox.

Our PIML workflow (**Figure 2**), which offers interpretable de novo network models, presents a paradigm shifts in terms of the investigative power of regulation analysis. With this workflow, we created two versions of de novo regulatory networks, the iModulon and Gene models, from the same curated dataset. The de novo networks are represented as directed graphs with connections between nodes, allowing interpretation of relational interactions from direct or indirect paths. Both iModulon and Gene models learned from curated datasets provide valuable insight into the interpretable outcomes of PIML, offering a fine- or coarse-grained view of the regulatory network at a hub (e.g. a group of genes such as a regulon) or at a gene scale, respectively, over a network architecture. The iModulon model revealed a multilayered web of interactions between iModulons while the Gene model revealed directed regulations between genes. Both models not only captured known interactions between TFs, but also inferred new connections to others in a de novo network. There are tradeoffs, however, in terms of the computational cost as the Gene model that provides a fine-grained view of the data structure increases training difficulty. Nevertheless, all simulations were completed less than a week on high-performance computing clusters.

## The de novo networks revealed multilayered regulations of central carbon metabolism of *S. elongatus* for the efficient energy utilization in a light-energy-limited environment

We constructed a data-driven workflow to create simplified *de novo* networks that recapitulate the regulation of carbon fixation and reveal the relational interactions among the genes or iModulons under light disturbance. The curated transcriptome datasets encompass regulatory gene networks and several overlapping enzymatic reactions within the central carbon metabolisms. These include light-dependent or -independent redox regulation of photosynthetic electron transport, the Calvin-Benson cycle, and starch biosynthesis (**Figure S4**).

The iModulon model of networks from PIML provides a global picture of a complex cascade of regulatory interactions, including a circadian rhythm that alternates the daytime and nighttime metabolic mechanisms for survival. Comparison of iModulon gene sets to previously published iModulon analysis in PCC 7942 (36) demonstrated conserved iModulons with high gene set similarity (3**, Table S11**). These iModulons identify regulatory signals in relation to gene sets enriched in transcriptional targets of global regulators RpaA, RpaB and sigma factor SigF2. Characterization of PIML learned interactions responsive to light perturbation demonstrates connections between RpaA (37, 38) and downstream sigma factors essential for controlling the metabolic switch from day to night (31).

In the daytime, the reductants produced from photosynthesis are consequently used for carbon fixation and glycogen synthesis via the Calvin-Benson-Bassham cycle. In darkness, cyanobacteria rely on the oxidative pentose phosphate pathway (OxPPP) (39) to produce reductants in the form of NADPH and downregulate DNA replication (40, 41) to conserve energy (39, 42) and to detoxify ROS when photosynthesis is not active. The coarse-grained de novo network (the iModulon model in **Figure 3A**) revealed a direct interaction from master circadian regular RpaA and sigma factor SigF2 to iModulons enriched in the regulatory targets of each transcription factor in response to light perturbations. These direct interactions identify experimentally characterized roles for RpaA and SigF2 in regulating the transition to night. iModulon gene sets correspond to the primary circadian regulon (RpaA), a gene set enriched in targets of RpaA and RpaB (RpaB), a gene set enriched in SigF2 targets containing genes involved in DNA uptake (Competence).

In contrast, the fine-grained de novo network presented by the "Gene" model revealed far more hidden interactions between genes than the known ones in **Figure 3B.** Here sigma factor RpoD5 downregulates almost every node on the graph with high confidence . Interestingly, the downregulated gene products are spread throughout the interior of the model cyanobacteria, including the inner membranes, cytosol, thylakoids, and DNAs. The inference from regulating genes to the spatial complexity of morphological phenotypes indicates a dynamic link between gene regulation and cell growth optimized by circadian rhythmicity in cyanobacteria adapting to light-energy limits from environments.

## Tracking dynamic redox reactions in response to light disturbance effectively integrates transcriptome and proteome datasets in cyanobacteria and leads to the discovery of protein regulators

We used the curated transcriptome datasets as a guide to develop a hypothesis about the proteins whose chemical activities are in action after light disturbance on a non-equilibrium energy landscape (**Figure 1A**). We hypothesized that the time required to change protein abundance would be longer than the time required for changes in protein redox states under light stimuli, due to the elaborate mechanisms involved in protein synthesis compared to the prompt cellular redox reactions. The former represents slow regulatory responses through protein-DNA binding that involves conformational changes such as protein folding and DNA bending dynamics. In contrast, the latter follows photo-induced electron transfer that suddenly changed a large amount of ROS and protein redox states.

We tested this hypothesis and compared the global and redox proteome for the gene products from the curated transcriptome datasets. After two hours in darkness from constant light exposure, we discovered a few protein products, namely Fbp, OpcA, RpaA, and ZWF, with significant changes in the cysteine oxidation level, while there were insignificant changes in their protein abundances (**Figure 5**). This signifies the assemblage of the four protein regulators operate in an immediate time scale involving structural dynamics of proteins with PTMs in biochemical reactions.

We highlighted the potential overlap and distinctions between light-driven regulations at the expression level and redox-dependent regulations at the protein structure level in *S. elongatus*. Fast redox reactions occur through cysteine PTMs that alter structural complexes and dynamics, offering a rapid and specific response to environmental stress from reactive oxygen species (ROS). To this end, tracking the redox reactions in response to light effectively integrate information from transcriptome and promptly screens for regulators in the redox proteome datasets. By combining TF action information with machine learning, our approach provides a novel set of tools to drive insights into complex cell responses and enable new discoveries in gene regulatory interactions.

## "Circadian rhythmicity interchange" assemblage connects an inbound photoinduced electron transfer highway and an outbound reduction-oxidation reaction highway to multiple phenotypes in the central carbon metabolism of *S. elongatus*

We speculated that the assemblage comprising of RpaA, Fbp, Zwf, and OpcA play a critical role in directing the upstream electron-energy transfer pathways and the downstream carbon metabolism pathways. Each gene produce comes from different pathways and together they form a functional construct similar to an "interchange" on the junction of road transport with busy traffic (**Figure 6**). This surprising discovery suggested, within the first two hours, that these four protein regulators respond to light disturbance through fast cysteine modifications on protein structures, rather than through gene regulations.

This interchange is regulated by cyanobacteria's circadian timed through RpaA. RpaA (regulator protein of phycobilisome association A) is a master circadian regulator of phycobilisome, a megacomplex on the thylakoid. De novo modeling infers that the regulatory protein of programmed circadian control (RpaA) indirectly coordinates the genes (*opcA*, *fbp* and *zwf*) encoding enzymes in the two alternating pathways for producing reductant NADPH during the day-night cycle. A gene product of *fbp* is Fructose-1,6-bisphosphatase class 1 (43, 44), known as a redox-sensitive protein in the Calvin-Benson cycle (CBC) whose activity can be regulated by light. A gene product of *zwf* is glucose-6-phosphate 1-dehydrogenase (G6PD). OpcA(45), an allosteric effector mediating the activity of G6PD, activates the pentose phosphate pathway when light-derived energy is not available [52]. A cysteine-rich regulatory protein, OpcA, is speculated to shunt biochemical reactions in the Calvin-Benson cycle and the oxidative pentose phosphate pathway (OxPPP) through cysteine PTMs.

In our redox proteome study after light disturbance, we revealed the opposing changes in cysteine oxidation activities on cysteines for the gene products of *fbp* and *zwf* (**Figure 5B**), while OpcA carried multiple redox shifts in either more reduced or more oxidized tendency. We speculated that RpaA, OpcA, ftp, and zwf form a "circadian rhythmicity traffic interchange" to transport electron-energy from inbound photo-induced reaction "highways" to outbound redox reaction "highways" (**Figure 6**). The output of metabolic states is directed by cyanobacteria's circadian, RpaA, that directs reductants from fast electron-energy reactions dependent to the availability of light to the central carbon metabolism via timed alternating pathways – CBC and OxPPP -- with overlapped biochemical reactions. Our work elucidated the assemblage of circadian rhythmicity interchange as a iModulon that interacts with other cellular processes such as photosynthetic light reactions, the electron transfer chain, and cellular metabolism via redox post-translational modification into a broader cellular context (46). We argued that the energy-landscape concept (16) provides the framework to describe complex reactions built on distinct time scales of chemical and molecular interactions in non-equilibrium cellular processes crucial for cyanobacteria's entrainment to the periodic light-energy limitations observed in diel cycles (39).

## Conclusion

Our workflow, enhanced with the PIML models, integrates transcriptome and proteome datasets in cyanobacteria to track dynamic redox reactions in response to light disturbance. We elucidated cellular response to perturbations through a nonlinear kinetic model of genetic expression without prior knowledge of the genes and transcription factor interactions. This flexibility allows the application of our approach to large datasets, enabling the construction of fine-grained or coarse-grained *de novo* regulatory networks with interpretable outcomes. The integrated framework of PIML and proteomics, PIML-omics, provides scientists with insights to develop testable hypotheses about the roles of gene products and protein regulators in dynamic cellular processes, driving new discoveries in gene regulatory interactions and new ways to integrate -omics experiments using illumination as disturbance. Overall, our approach is a paradigm shift that offers a robust and adaptable tool for advancing our understanding of cellular responses, with broad applicability for foundational biological physics research and bioengineering applications.

## Materials and Methods

### Collection of *S. elongatus* transcriptome datasets

The training datasets were processed from an existing transcriptome database of *S. elongatus*. Raw RNA sequencing (RNA-Seq) data as FASTQ files from the NCBI Sequence Read Archive (SRA, (47)), Gene Expression Omnibus (GEO, (48)) and Joint Genome Institute databases (JGI, (48)) (as of 31 January 2023) were collated into a single training dataset. Overall, the raw RNA-Seq dataset comprised of 386 samples as described in **Table S1**. Raw sequences were processed using Rsubread (49) and mapped against a set of reference genome sequences: chromosome NC 007604.1, pANL plasmid NC 004073.2, and pANS plasmid NC 004990.1. The resulting mapped RNA-Seq dataset was assessed using FastQC (50) and manually inspected and refined by down selecting only data with a high correlation between biological replicates (>0.9) as well as the associated metadata for the experimental conditions, as described provided in the sourced databases or related peer-reviewed publications (25, 28, 51-56). In total, the resulting RNA-Seq dataset for PCC 7942 comprised of 330 high-quality RNA-Seq samples (named selongEXPRESS) and is described in **Table S2**.

### Curation of modulated datasets with ICA

We transferred the full gene dataset to a modulated data set where the dimensions of the input gene dataset were reduced using ICA on the iModulonDB (9) web platform. The log-transformed transcripts per million (TPM) gene count samples that passed the QC pipeline were used as input for ICA (S3 Table). Then, ICA was performed to calculate the independent M and A significance matrices (S4 and S5 Tables), which correspond to the contribution of each gene to the statistical independent source signal in the iModulons and the iModulon signal strength for a given condition respectively. The reduced model consists of a set of "perturbations", "molecular transcription factors", and "phenotype iModulon responses". We explained these terms in the Supplementary Information (S3. Curation of modulated datasets with ICA).

### Construction and interpretation of a reduced regulatory network from training system-level PIML models with the CellBox package

We utilized the CellBox(24) package to predict the responses of each TF, gene, or iModulon as a node in an interaction network under perturbation conditions. These interactions in response to perturbations (e.g.

illumination) were generated by solving the coupled ordinary differential equation (ODE) system in **Eq 1** with the long-time steady-state expression level that offers the predictions. The mathematical framework of the CellBox package is the approach of perturbation biology for inferring network models(22) in cellular systems described below.

$$\frac{dg_i(t)}{dt} = \epsilon_i \tanh\left(\sum_{j\neq i} \omega_{ij} g_i(t) + p_i(t)\right) - \alpha_i g_i(t) \qquad \textbf{Eq (1)}$$

It includes an independent time variable $t$, which denotes that variables such as expression level, $g$, and perturbation, $p$, are functions of $t$. The connectivity interactions are described by $\omega$ and the magnitude of these interactions, $\epsilon$. The parameter describing the decay of the expression level is provided by $\alpha$. Sander *et al* (57) developed a statistical physics approach to computationally solve the coupled ordinary equations (ODEs) from a large solution space of all possible network models using a probabilistic algorithm, Belief Propagation (BP), which is three orders of magnitude faster than standard Monte Carlo methods. Later, they deployed a data-driven machine learning workflow in the Cellbox package to ensure every expression level reaches a steady state. Here in our study, we leveraged Cellbox to model the expression levels of TFs and genes (or iModulons) in response to a light perturbation over the natural clock of cyanobacteria where the interactions of expressions (or regulations) are used to infer their contributions towards a steady state expression.

A full description of the equation and simulation input/output using Cellbox is provided in **Section S6** of the Supplementary Information. The light intensity and photon flux/cycle time metadata were encoded into system perturbations, as photons per micromole per second, which allow us to explore the effects of light on the cell system. The best models returned from the CellBox package were analyzed to determine the inferred interaction networks. The significant gene or iModulon interactions were defined from the output network by using an FDR adjusted t-test, in which edge interactions were tested for a significant difference, with edges with a p-value greater than 0.023 (or 0.015) being retained for the iModulon (or Gene) model, respectively.

Taking the best models trained by CellBox, networks were then represented as graphs to facilitate the interpretability of the outcomes for both the iModulon and Gene models. This network of a reduced regulatory system demonstrates how the pairwise interactions (58) advance the understanding of the causal interactions within the corresponding curated datasets. We used the values within the parameter matrix $\omega$ in **Eq 1** that contain the interaction information to capture the influence from the other genes or iModulons on a given gene. These values can be used to draw an de novo interaction network on a graph where the nodes are the set of genes or iModulons, and the edges show the interactions. The network interactions for both the iModulon and Gene models were analyzed to highlight the significant interactions justified by the one-sided t-test which queries the significance in the learnt interaction strengths when compared to the null hypothesis of no interaction.

## Global and redox proteome experiments

Global and redox proteome experiments were performed to understand the regulatory dynamics of carbon fixation in *S. elongatus* PCC 7942 cscB/SPS. This microbe was cultured in a modified BG-11 medium 0.09 g $L^{-1}$ Yeast Nitrogen Base without amino acids and ammonium sulfate (H26271.36, Thermo Fisher Scientific Inc., USA), 0.264 g $L^{-1}$ of $(NH_4)_2SO_4$ (J64180.A1, Thermo Fisher Scientific Inc., USA), 0.174 g $L^{-1}$ of $K_2HPO_4$, (60,356, Sigma-Aldrich, USA), and 1 mM of isopropyl-β-d-thiogalactopyranoside (I56000, RPI, USA) at 30°C with a 1% $CO_2$ supply. The culture was exposed to constant light (~650 μmol $m^{-2}$ $s^{-1}$) for 6 days then diluted with spent culture media to an OD750nm of 0.08 (to enable full and equal light availability in the culture) and cultured under lights for 30 min.

Three biological replicates for each treatment were prepared as follows. The light treatment samples were harvested following the 30 min of light exposure, while the dark treatment cultures were transitioned to the dark (light turned off, culture flasks wrapped in foil) and incubated for an additional 2 hr before collection and sample processing. For each biological replicate, 100 mL of culture was split into two 50 mL aliquots and centrifuged at 4,700 g for 5 minutes at 4˚C. The sample supernatant was decanted, and the cell pellets were flash frozen in liquid nitrogen and stored at –80˚C. Dark treatment samples were kept in the dark during these processing steps. Next, samples were processed as described previously (33). Briefly, cell pellets were treated with 10% cold trichloroacetic acid on ice. Then, cells were incubated in 250 mM HEPES (pH 7.5) containing 8 M urea, 10 mM ethylenediaminetetraacetic acid (EDTA), 0.5% sodium dodecyl sulfate (SDS), and 100 mM N-ethylmaleimide (NEM) for 2 h, then lysed by bead beating. The extracted protein was precipitated and washed with cold acetone, then reduced with dithiothreitol. The reduced protein was further processed with a resin-assisted capture (RAC) protocol to selectively enrich proteins with oxidized cysteine residues (59). The detailed experimental procedure for redox proteome is further described elsewhere (60).

At the same time, an aliquot was taken from each protein sample before resin-assisted capture enrichment to quantify the global protein expression level. Isobaric labeling was used to assist quantification. The enriched samples and global protein samples were analyzed by a nanoAcquity UPLC system (Waters) coupled to a Q-Exactive HF-X Orbitrap mass spectrometer (Thermo Scientific, San Jose,

CA) in 120-min liquid chromatography (LC) gradient. A full mass spectroscopy (MS) scan was acquired in the range of m/z = 400 – 1800. MS/MS was performed in a data-dependent mode. The parent ion tolerance was 20 ppm. LC-MS/MS raw data were searched against the *S. elongatus* PCC 7942 UniProt database (downloaded on March 8, 2022) using MS-GF+ (59, 61). The detailed experimental procedure is further described elsewhere (59).

## Word cloud analysis

The proteomic results were filtered by a spectral-level FDR of less than 1%, based on a target-decoy strategy and MSGFDB_SpecEValue < $1x10^{-8}$. Data analysis was performed using R. The expression measurements were analyzed for significance. Values with a log2 fold change > 0.1 or < -0.1 with a p-value < 0.05 were considered significant. These genes were cross-referenced with the sets of significant genes returned by the computational analyses to highlight mechanical importance, with results from the computational workflow from this study. Word clouds were formed by comparing the significant changes in gene expression between the PIML models and in the proteome expression data.

## Acknowledgments

The research described in this paper is part of the Predictive Phenomics Initiative and was conducted under the Laboratory Directed Research and Development Program at Pacific Northwest National Laboratory. This work is also partially supported by the NW-BRaVE for Biopreparedness project funded by the U. S. Department of Energy (DOE), Office of Science, Office of Biological and Environmental Research, under FWP 81832. A portion of this research was performed on a project award (https://doi.org/10.46936/staf.proj.2023.61054/60012367) from the Environmental Molecular Sciences Laboratory, a DOE Office of Science User Facility sponsored by the Biological and Environmental Research program under Contract No. DE-AC05-76RL01830. Pacific Northwest National Laboratory is a multi-program national laboratory operated by Battelle for the DOE under Contract DE-AC05-76RLO 1830.

**References**

# Figures

## Figure 1.

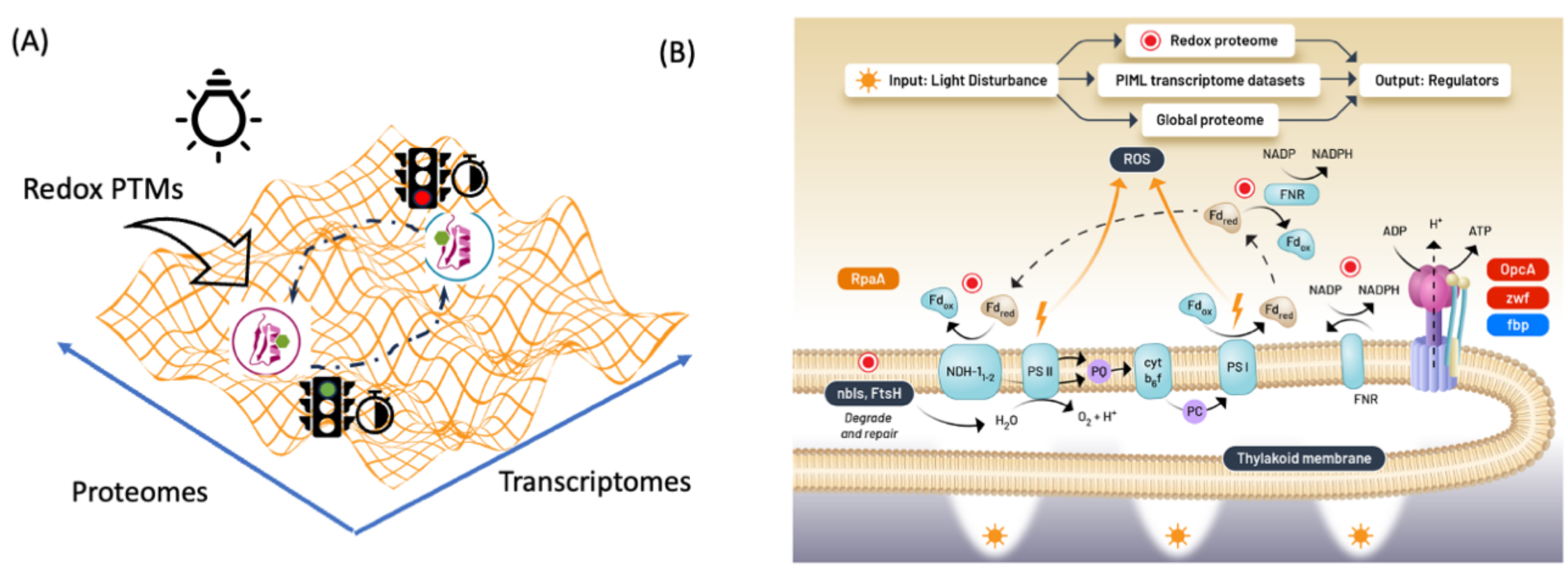

**A general framework for reaction dynamics in complex system and paths for an out-of-equilibrium process of electron and energy transport over the diel cycle in *S. elongatus* for carbon fixation. (A)** A functional state on the landscape characterized by proteome and transcriptome is represented by the activities of protein structures. Cysteine PTMs on protein structures signal the temporal alternation of the metabolic pathways in phase with the periodic light-energy limitation from the environment. New phenomena, or new metabolic states, can arise from the mesoscopic dynamics of intrabasin or interbasin motion on a non-equilibrium energy landscape under light disturbance. (B) Model cyanobacteria *S. elongatus* serves as a science use case (25-28). During the day, energy is generated via photosynthesis, with several proteins involved in the transport of electrons and the reduction of electron donors. The photosynthetic electron transport chain begins with photosystem II (PSII) on thylakoids. Cyclic electron flow is another mechanism of electron-energy transport where the cyanobacteria handle stress from light exposure by redirecting electrons from PSI to NDH-1 through ferredoxin (Fd) and plastoquinone (PQ) and splitting $H_2O$. In our study, we discovered four protein regulators as outputs from the workflow, shown in colored boxes, that collectively responds to cysteine post-translational modifications within two hours of light disturbance from the *S. elongatus* culture, while their relative protein abundance remains similar. The orange box represents RpaA, a master regulator from the circadian cycle. The red boxes represent OpcA and zwf from the oxidative pentose phosphate pathways. The blue box represents fbp from the Calvin-Benson cycle.

**Figure 2.**

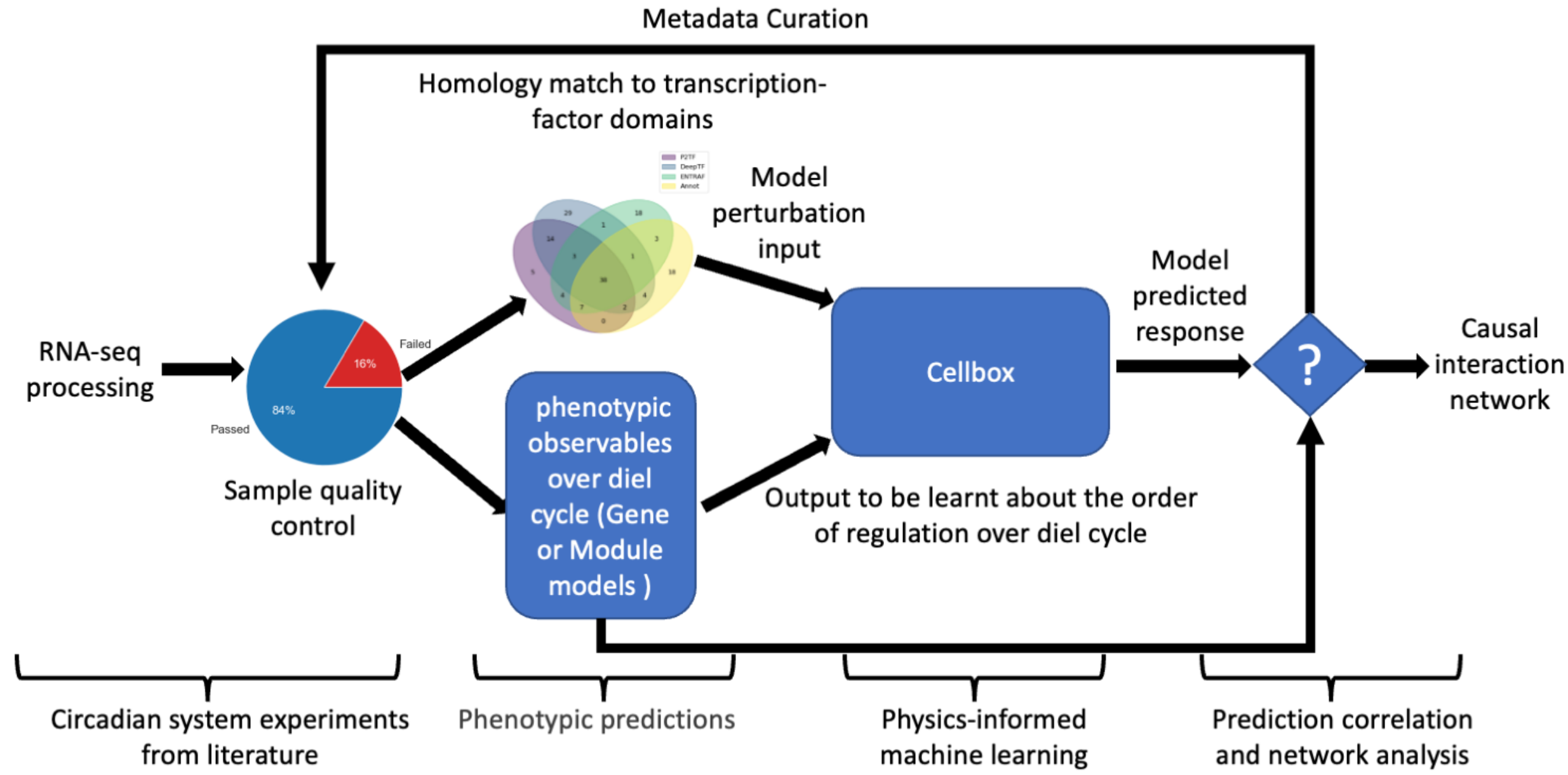

**Our iterative workflow for creating a transcriptomic dataset of metadata related to S. elongatus.** The overall workflow involves iterative steps to create a curated dataset of metadata from the transcriptome in response to light stimuli. We used the gene expression as a proxy for phenotypic observables, grouped in Gene or iModulon models, over the diel cycle to construct a mathematical representation of dynamical equations using a physics-informed machine-learning (PIML) approach.

**Figure 3.**

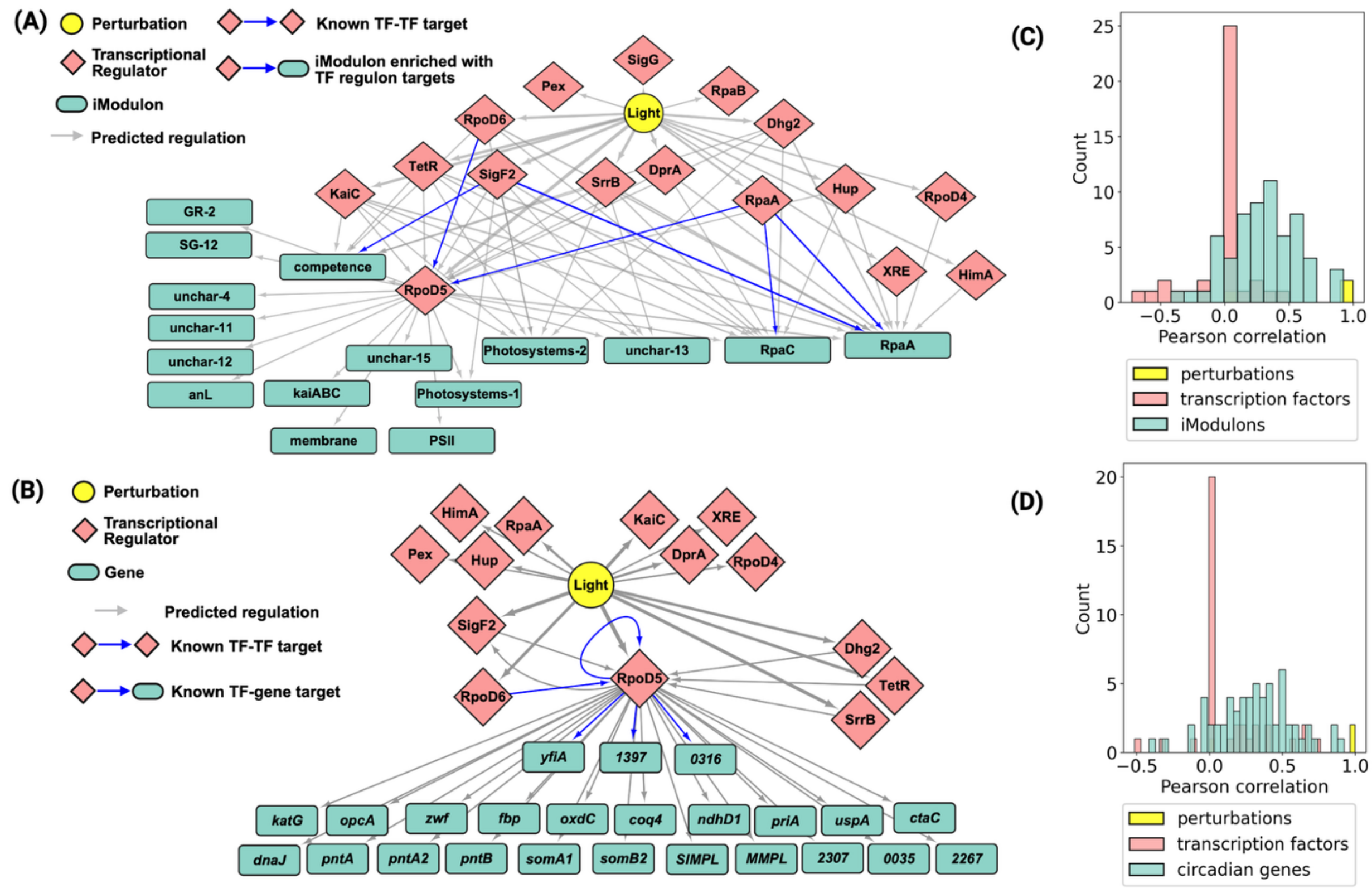

**Comparison of gene interactions for the iModulon and Gene models.** The interactions between entities in the (A) iModulon model and (B) in the Gene models are revealed by a directed graph from a simplified regulator network responsive to light stimuli. (C, D) Histograms of the node counts sorted by the Pearson correlation coefficients between the predicted and observed expression levels for each node for transcription factors (TFs) or perturbations. The descriptions of the labels are provided in **Table S13, Table S14** and **Table S15.**

**Figure 4**

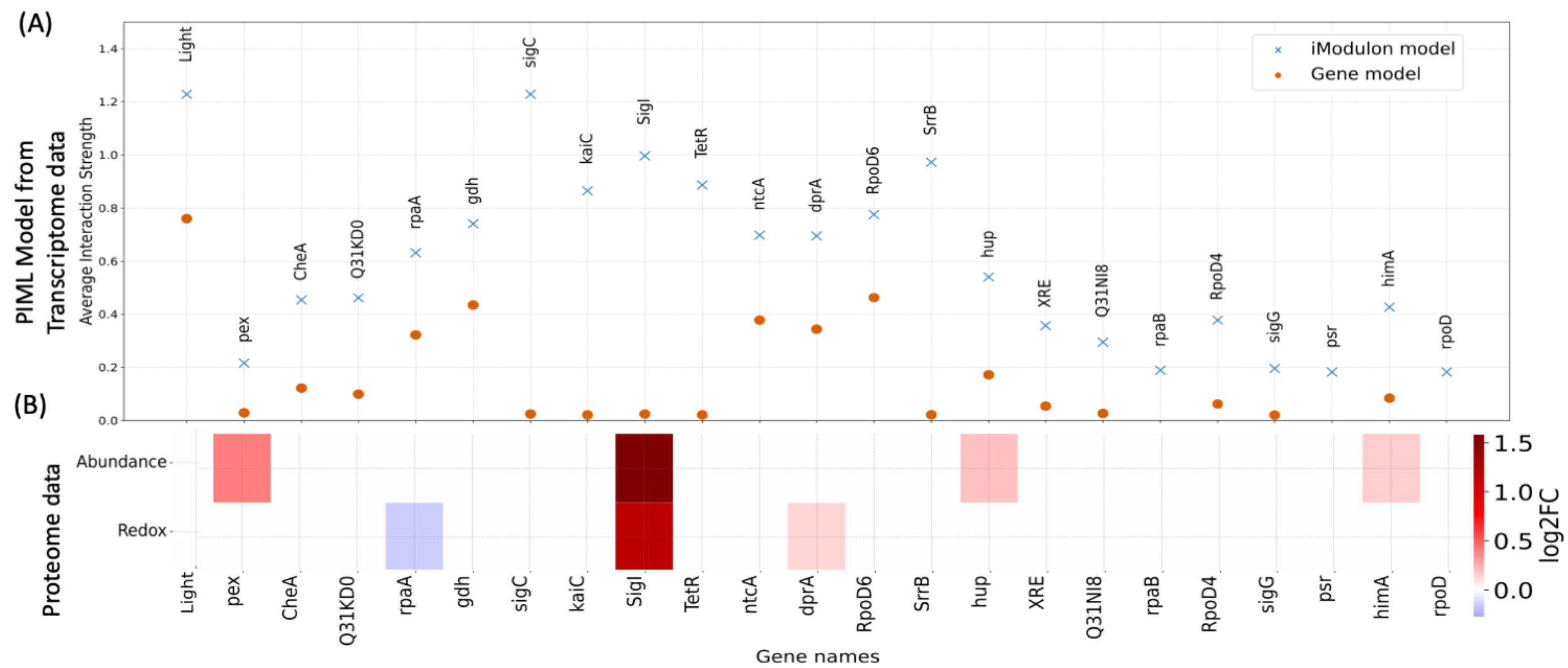

**Testable predictions from de novo kinetic models based on gene expression data for transcription factors in conjunction with redox proteome analysis under light disturbance**. (A) The magnitude of inferred interaction strengths meeting t-test significance criteria that contribute towards a steady-state response to the light perturbations for the transcription factors (TF) in the de novo kinetic models. For the iModule model (blue cross), TFs sigC (RpoD5), SigI and SrrB drive the response to light perturbation, while for the Gene model (red dots), gdh, RpoD6, and ntcA are strong drivers. (B) The log2FC changes in the protein abundance expression from the global proteome as well as the redox expression from the redox proteome of *S. Elongatu*s under illumination perturbation. The plot shows protein abundance changes for pex, hup, and himA, but not their redox expression. Redox expression changes for rpaA and dprA, but not their abundance expression.  Both protein abundance and redox expression change for SigI.

**Figure 5**

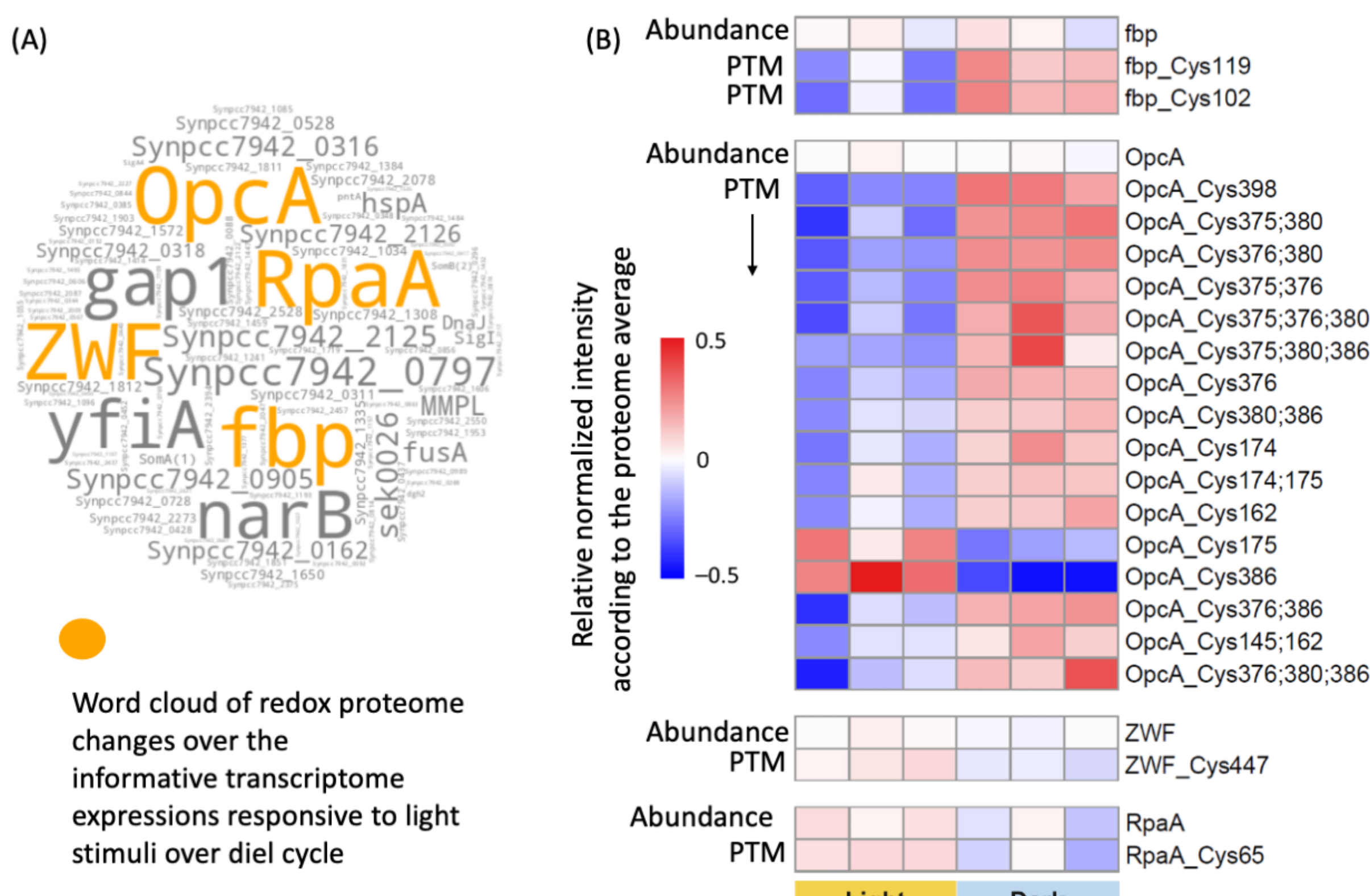

**Identified proteins with redox post-translational modification (PTM) expressions that changed significantly under light/dark perturbations.** (A) A word cloud was used to gain mechanistic insights into the significance of gene responses under light/dark perturbations. Gene expression responses to light stimuli are derived from the curated dataset of metadata, with those capturing the significant changes in the protein PTMs as revealed by the redox proteomics experiment colored in orange. (B) The heat map of normalized intensity from two kinds of proteome experiments illustrates the variations in the observations. The blue to red scale represents normalized intensities of relative protein abundance (i.e. 1st row of each protein heatmap) and cysteine oxidation level (i.e., cysteine PTM, from the second row to the last of each block). The normalized intensity was scaled by log2. The left three columns of the heatmaps are the replicated measurements under light conditions and the right three columns are measurements under dark conditions taken two hours after switching from constant light exposure.

**Figure 6.**

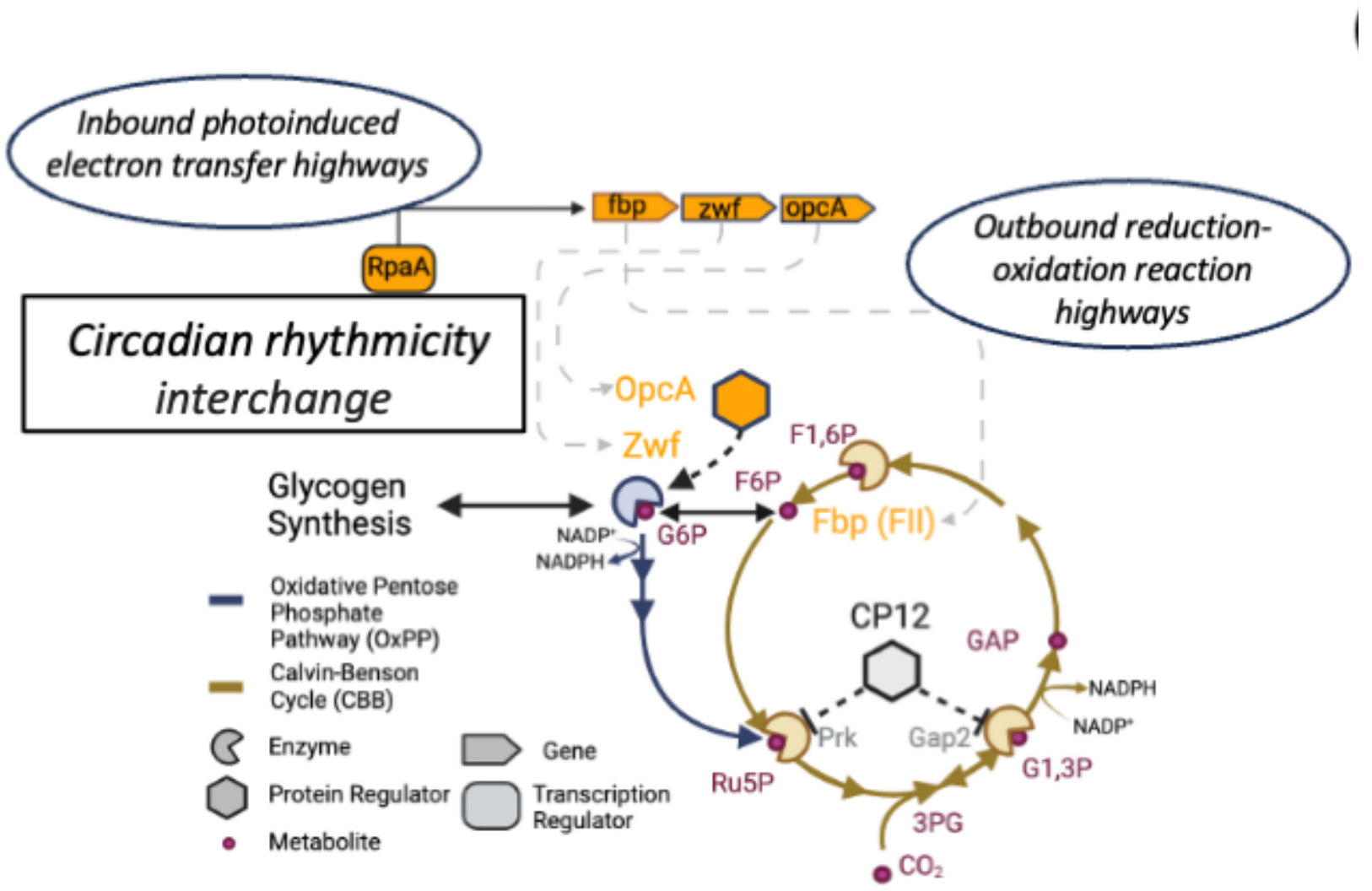

Programmed **circadian rhythmicity interchange in *S. elongatus*.** A conceptual map of circadian rhythmicity *interchange* connects an inbound photoinduced electron transfer highway and an outbound reduction-oxidation reaction highway to multiple phenotypes in the central carbon metabolism of *S. elongatus*. The assemblage of *interchange* comprises RpaA, fbp, zwf, and opcA (shown in orange). The redox of cysteine PTMs on these gene products, or proteins, signals and coordinates the inbound electron transfer pathways and outbound central carbon metabolism responsive to light disturbance. A cysteine-rich regulatory protein, OpcA, is speculated to biochemical reactions in the Calvin-Benson cycle and the oxidative pentose phosphate pathway (OxPPP) through cysteine PTMs. Our study infers that the regulatory protein of circadian control (RpaA) indirectly coordinates the genes (opcA, fpb and zwf) encoding enzymes in the two alternating pathways for producing reductant NADPH during the day-night cycle.

**Supporting Spreadsheet with Tables from S1 to S10.**

**Table S1. Overview of the raw RNA sequencing data .**
**Table S2. High-quality RNA-seq dataset, "selongEXPRESS", that passed the quality control.**
**Table S3. Transformed gene counts for the selongEXPRESS dataset that passed the quality control.**
**Table S4. Calculated independent A significance matrix.**
**Table S5. Calculated independent M significance matrix.**
**Table S6. Gene expression log2 fold changes calculated using DESeq2.**
**Table S7. Standard error values for the log2 fold changes calculated using DESeq2.**
**Table S8. Adjusted p-values for the log2 fold changes calculated using DESeq2.**
**Table S9. ICA results**
**Table S10. List of circadian genes**

**Supplementary information with Tables from S11 to S15 and Figures from S1 to S5**

**Table S11. List of known transcription factors responsible for circadian rhythm adjustments from the literature.**
**Table S12: Summary of the statistics of the mathematical models trained by CellBox using the best curated datasets from the transcriptome**.
**Table S13. iModulon homology and naming versus published results**
**Table S14. Descriptions of the node labels on the de novo network.**
**Table S15. Descriptions of the additional labels on the de novo network.**

**Figure S1. Correlation between the experimentally measured values from the literature and the predicted response values from CellBox**
**Figure S2. Grouped iModulons from the curated metadata**
**Figure S3. iModulon gene set homology versus published iModulon analysis**
**Figure S4. Overlapped enzymatic reactions within the central carbon metabolism of PCC7942.**
**Figure S5. Testable predictions from de novo kinetic models based on gene expression data for genes found in RpaA iModulon in conjunction with redox proteome analysis under light disturbance.**
**Figure S6. Testable predictions from de novo kinetic models based on gene expression data for genes found in competence iModulon in conjunction with redox proteome analysis under light disturbance.**